\documentclass[runningheads]{llncs}
\usepackage[T1]{fontenc}
\usepackage{graphicx}

\usepackage[T1]{fontenc}
\usepackage{graphicx}
\usepackage{stmaryrd}
\usepackage{amsfonts,amstext,amsmath, amssymb}
\usepackage{mathrsfs,stmaryrd}
\usepackage[linesnumberedhidden]{algorithm2e}
\usepackage{todonotes}
\usepackage{tikz}
\usepackage{extarrows}
\usepackage{mathtools}
\usepackage{hyperref}
\usepackage{multicol}
\usepackage{cleveref}
\usepackage{enumitem}
\usetikzlibrary{calc,shapes.multipart,chains,arrows}
\usetikzlibrary{decorations.pathreplacing}

\usepackage{listings}
\definecolor{codegreen}{rgb}{0,0.6,0}
\definecolor{codegray}{rgb}{0.5,0.5,0.5}
\definecolor{codepurple}{rgb}{0.58,0,0.82}
\definecolor{backcolour}{rgb}{0.95,0.95,0.92}

\usepackage[font=small,skip=2pt]{caption}

\newcommand{\sem}[1]{\ensuremath{L(#1)}}
\newcommand{\set}[1]{\{#1\}}

\newcommand{\Nat}{\mathbb{N}}

\newcommand{\ksla}{k\textit{-SLA}}

\newcommand{\emptylookback}{\varepsilon}

\newcommand{\trans}[1]{\xrightarrow{#1}}

\newcommand{\pa}{\mathit{\textcolor{red!50!black}{PA}}}
\newcommand{\wa}{\mathit{\textcolor{green!30!black}{WA}}}
\newcommand{\startNewPane}{\mathit{startNewPane}}
\newcommand{\prefState}{\ensuremath{\mathit{PrefixState}}}
\newcommand{\windowStartIndices}{\ensuremath{\mathit{WindowStartIndices}}}

\newcommand{\pluseq}{\mathrel{+}=}

\newcommand{\initwa}{\ensuremath{\textit{\textcolor{green!30!black}{init}}_\wa}}
\newcommand{\initpa}{\ensuremath{\textit{\textcolor{red!50!black}{init}}_\pa}}
\newcommand{\wordblock}{\ensuremath{\textit{w\_block}}}
\newcommand{\sre}{\ensuremath{\mathit{SRE}}}

\newcommand{\lang}[1]{\ensuremath{L(#1)}}

\newcommand{\val}{v}
\newcommand{\restr}{\upharpoonright}
\newcommand{\initS}{\ensuremath{\textit{init}_S}}

\newcommand{\para}[1]{\smallskip\noindent\textbf{#1.}}

\begin{document}

\title{WEX: Formal Specifications for Windows in Stream Processing\thanks{M. Praveen was partially support by the Infosys foundation}}

\author{S Hitarth\inst{1}\orcidID{0000-0001-7419-3560} \and
M. Praveen\inst{2,3}\orcidID{0000-0002-5734-7115}
}
\authorrunning{S Hitarth and M. Praveen}
\institute{Hong Kong University of Science and Technology, Hong Kong \and
Chennai Mathematical Institute, India \and
CNRS IRL ReLaX, Chennai, India,\\
\email{hsinghab@connect.ust.hk, praveenm@cmi.ac.in}}

\maketitle              %

\begin{abstract}
A key operation in processing an unbounded data stream is windowing, which
extracts finite portions of streams for further handling.
The existing frameworks and query languages either require windows to be
defined using ad hoc imperative languages or are limited to rudimentary
constructs such as time- or count-based
windows. 
We propose \textbf{W}indow \textbf{EX}pression, a formal specification for precisely expressing 
windowing constructs
based on monadic second-order logic. WEX can naturally express traditional windowing constructs such
as sliding windows and tumbling windows, as well as more complex windows whose start
and end indices are triggered based on the satisfaction of given logical conditions.
After introducing a model of 
\textit{symbolic 
automata with lookbacks} over an alphabet theory, we present another equivalent
representation of WEX based on symbolic regular expressions. The precise semantics
of windowing enable static analysis over WEX. In particular, we show that, in general,
it is undecidable to check whether a WEX allows an unbounded number of overlapping windows.
However, when the data stream is over a finite alphabet, or the alphabet theory has 
the so-called
\textit{completion property}, the problem becomes decidable.

\end{abstract}

\section{Introduction}
\label{sec:intro}

\para{Stream Processors} 
Traditional ways of storing and querying data do not work well in
scenarios where data is being generated continuously and quick
decisions need to be taken. For example, in hospital intensive care units, signals from multiple devices need to be monitored and the occurrence of
any anomaly should raise alarms immediately.
Stream Processors are programs that consume and produce  such
unbounded streams of data. 
They are applied in many areas, ranging from
detecting who is controlling the ball in soccer matches \cite{PS2015} to
detecting irregularities in heartbeat rhythms in implantable
cardioverter-defibrillators (ICDs) \cite{ARMBSG2019} and continuous 
analysis
of RFID readings to track valid paths of shipments in inventory
management systems \cite{WL2005}. One common aspect is that they 
produce output within a bounded amount 
of 
time, during which they can only read a bounded portion of the input. 

\para{Windows} Windows define a span of positions along a stream that the program 
can use as a unit to perform computations on and are fundamental to 
stream processors. 
Stream processors allow the end-users to specify windows using their specification language, however, not all of them allow end-users to define 
customized windows.

\begin{example}\label{example:sliding-windows}
Suppose we want to generate the so-called \textit{sliding windows} of size $5$ and offset $2$, i.e., each window will span $5$ data elements and the next window will be generated by sliding the current window by $2$ data elements. This is illustrated in Figure \ref{fig:ill-windowing}.

\begin{figure}[htb]
\begin{center}
\begin{tikzpicture}[shorten >=2pt,->]
\def\recsize{0.5}
  \tikzstyle{vertex}=[rectangle,fill=blue!5, minimum size=12pt, inner sep=0pt, minimum width = \recsize cm, minimum height = \recsize cm, opacity=0.9]
  \tikzstyle{enclose}=[rectangle, thick, draw=blue!90, minimum size=12pt, inner sep=0pt, minimum height = \recsize cm, opacity=0.9]

 \foreach \text/\name/\x in {a_0/a_0/0, a_1/a_1/1, a_2/a_2/2, a_3/a_3/3, a_4/a_4/4, a_5/a_5/5, a_6/a_6/6, a_7/a_7/7, a_8/a_8/8, a_9/a_9/9, {a_{10}}/{a_10}/10, {a_{11}}/{a_11}/11, \ldots/{a_{12}}/12}
        {
        \node[vertex] (G-\name) at (\x/2,0) {$\text$};
        };

  \foreach \wtext/\windowstart/\windowsize\voffset in {2/0/5/0, 3/2/5/-0.08}
  {

        \pgfmathsetmacro\winw{(\windowsize)/2)};
        \pgfmathsetmacro\wshift{(\windowstart*\recsize)+(\winw*\recsize)-(\recsize/2)};
        \ifthenelse{\windowsize > 0}
          {\node[enclose, minimum width=\winw cm, thick, dashed, draw=orange!100, opacity=0.9] (\wtext) at (\wshift,\voffset) {}}
          {};
   }

    \foreach \xstart/\yend/\ystart/\text in {
    0/0/0.7/begin~$1$\;\;\;/, 
    2/1/0.7/begin~$2$/,
    5/2/0.7/end~$1$/,
    7/3/0.7/end~$2$/}
    {
        \pgfmathsetmacro\xpos{\xstart/2 - \recsize/2};      
        \pgfmathsetmacro\yendpos{((\recsize/2) - (\yend*\recsize))};
        \draw[draw=red!70, thick, opacity=0.9] (\xpos, \ystart) node[above, text centered, minimum height=0.5cm,  ]{\text} -> (\xpos, \recsize/2);
    }

\end{tikzpicture}
\end{center}
\caption{Illustration of Windows: Two windows $a_0a_1a_2a_3a_4$ and $a_2a_3a_4a_5a_6$ are generated}
\label{fig:ill-windowing}
\end{figure}

\end{example}
The survey \cite[Table IV]{CM2012} 
mentions thirty-four processors, of which six allow user-defined windows. Four of them 
allow only time-based or count-based windows. The other two (Esper 
\cite{Esper} and IBM System S \cite{IBMSystemS}) allow 
customizable windows based on other criteria, but they must be 
written in imperative programming languages.

The syntax of many 
languages used in stream processors extend database query 
languages. Relational database query processing arguably derives
a significant portion of its robustness from the fact that it is based on 
relational algebra that is expressively equivalent to first-order logic 
\cite{CoddsTheorem}. But for streaming processing, the fundamental linear 
order of arrival is not part of the syntax, which may be 
important for certain applications.

\para{Window EXpressions} We introduce a novel method of defining windows called \textbf{WEX} based on 
the Monadic Second Order (MSO) logic of one successor. MSO 
includes the linear order of arrival as a basic building block.  
Apart from easily expressing practical queries, we get other advantages from 
concepts and constructs in formal language theory. The equivalence of MSO
with regular expressions allows us to design another equivalent 
representation for defining 
windows that is arguably easier to understand for end users compared to 
logic-based syntax \cite{RiseAndFallOfLTL}. 
In particular, we give two equivalent definitions of WEX, one based on MSO and another based on regular expressions. The equivalence with automata 
allows us to design a procedure that automatically produces windows 
from a data stream according to specifications.

\begin{example}\label{example:sliding-window-with-mso}

Informally, WEX specifies windows with a
logical formula $\phi(x_s, x_e)$ over two special
free variables $x_s$ (start) and $x_e$ (end), such that 
the span $(a,b)$ of a word $w$ is a window if $w \models \phi(a, b)$.

For instance, the sliding windows from \Cref{example:sliding-windows} can be 
\textit{captured} by the
following formula in the theory of linear arithmetic:

$$\phi(x_s ,x_e) \coloneq \exists n \in \mathbb{N}, x_s = 2\cdot n \land x_e = x_s + 5.$$

\end{example}

\para{Motivating Example}\label{sec:motivating}
Sequence of stock prices as they 
are traded in the market is a natural data stream. 
Detecting trends in such streams is widely used in 
both stock markets and algorithmic trading \cite{PS2015}. 
Assume that the average price of a stock
is emitted every minute as a stream. 

\begin{figure}[h]
\centering
\begin{tikzpicture}[scale=0.45]

\draw[->] (0,0) -- (11,0) node[right] {Time};
\draw[->] (0,0) -- (0,6) node[above] {Average Stock Price};

\draw[thick,blue] plot coordinates {
  (0,3) (1,2.5) (2,2) (3,3) (4,3.5) (5,3.2) (6,4) (7,4.5) (8,4.2) (9,3.9) (10,4.2)
} node[right,black] {Price stream};

\foreach \x/\y in {0/3,1/2.5,2/2,3/3,4/3.5,5/3.2,6/4,7/4.5,8/4.2,9/3.9,10/4.2}
  \fill[black] (\x,\y) circle (3pt);

\draw[opacity=0.9,thick, dashed, draw=orange!100] (2,1.8) rectangle (9,5);

\draw[decorate,decoration={brace,amplitude=8pt,mirror}, thick]
  (2,1.5) -- (9,1.5) node[midway,below,yshift=-5pt]{Window};

\end{tikzpicture}
\caption{Stock price stream with average values emitted every minute. 
A window (highlighted) is used to detect a potentially downward trend.}
\label{figure:stock-trend-example}
\end{figure}

We are interested in detecting \textit{a trend} with the following behaviour,
illustrated in~\Cref{figure:stock-trend-example}. The window opens right
after the price starts to increase after having gone down at least twice 
consecutively. Inside the window, the price may fluctuate, but is not allowed
to go down twice in a row. The window closes right when the price
goes down twice consecutively. Such a window captures
a phase that is mostly increasing with occasional dips, ending with 
a potential downward trend. 

This kind of pattern cannot be specified by simple operators such
as sliding or tumbling windows as they are purely time or count-based. 
Nor do existing algebraic or stream-monitoring frameworks support 
such event-based windows defined rigorously as first-class objects:
they typically require imperative definitions of windows in 
some ad hoc language. 
In contrast, our formalism (WEX) can precisely define such intricate
patterns. A key motivation for WEX is to formalize such informal requirements,
which are otherwise prone to misunderstanding. We will formalize this with WEX
in \Cref{sec:examples}.

\para{Overlapping Windows} It is possible that a window specification 
results in a large number of 
windows overlapping at the same position of a data stream, 
overwhelming the stream processor. 
This is usually handled by engines using load shedding --- dropping off 
information items from the stream at runtime when load becomes too 
high. This may be acceptable if it is caused by high input rate, but not 
if there is a design fault in the window definition. We 
study the crucial problem of checking whether the 
number of windows overlapping at a single position is potentially unbounded for a 
given window definition. This problem is undecidable in general 
for infinite alphabets. We show that this is decidable when the
alphabet is finite and also when the alphabet theory has the so-called \textit{completion property}.

\para{Data Streams and Symbolic MSO} Data elements in a stream are usually numerical or similar values from an infinite domain. Symbolic MSO
\cite{SymbolicMSO} is meant to deal with infinite alphabets. The atomic 
formulas of this logic can check the properties of input symbols using 
predicates over the infinite alphabet. 
This also has automata counterparts, called symbolic automata 
\cite{SymbolicAutomata} and symbolic regular expressions 
\cite{SymbolicRegEx}. We will define these terms in the next section. Our window expressions can be expressed using a guarded variant Symbolic MSO. The advantage of high expressiveness of Symbolic MSO outweigh the disadvantage of undecidability of many static analysis problems over Symbolic MSO. Moreover, we identify a fragment for which it is decidable, even for symbolic automata.

\para{Contributions and structure} The main contributions of this paper are:
\begin{itemize}
    \item In \Cref{sec:prelim}, we introduce a variant of symbolic automata with 
    lookbacks ($\ksla$). 
    We estabilish certain closure properties of $\ksla$ to define symbolic regular expressions
    equivalent to it. We then prove that our automata model is equivalent to
    the Symbolic Monadic Second Order logic based on \cite{SymbolicMSO}.
    \item In \Cref{sec:WindowsSMSO}, we define the two formalism of WEX based on 
    a guarded fragment of S-MSO 
    and symbolic regular expressions. 
    In Section \ref{sec:examples}, we provide a few applications 
 of WEX to specify complicated windowing.
     \item We prove the decidability of the unboundedness 
 of overlapping windows for restricted classes of window 
 expressions in Section \ref{section:window-overlap-problem}.
    \item Finally, we present an algorithm for a stream processor based on WEX
    in Section \ref{sec:processor} that uses \textit{panes} to avoid duplicate
    computations.

\end{itemize}

\para{Related works} 
Mamouras et. al. extended regular expressions in
\cite{MRAIK2017} with operators to handle quantitative data for
processing data streams. In their work, windows have to be defined as derived operators or
written in external code, in contrast to our formalisms where 
windows are treated as first-class
objects and defining them is a basic construct.
A model called data
transducers is used in \cite{AMS2019} for implementing data stream
processors, but again windows are not part of the core specification
language. 

In runtime monitoring frameworks Lola/RTLola 
\cite{DAngelo2005LOLARM,faymonville2016stream-lola,baumeister2020rtlolaclearedtakeoffmonitoring}, one can define 
simple windows such as tumbling windows and sliding windows. 
In \cite{GRU2019}, a formal framework based on models of 
computation is developed for complex
event processing. They also work on data streams, but windows as we 
consider here are not of particular interest there. 
Algebraic
systems for generating windows from streams are considered in
\cite{PS2006,PLR2010}, and they study properties of window definitions
that are useful for query optimization and related static analysis
tasks. The windows are defined via algebraic constraints over 
the timestamps
of the streaming data, and not the actual values of the data. 
Moreover,
their goal is not to integrate window definitions in the syntax
of query languages, which we do here. 

In database theory, \textit{spanners} are used for information extraction 
from a document. Spans, like windows, are contiguous substrings of
the document identified by the starting and ending indices.
Formal spanners introduced in \cite{faginDocumentSpannersFormal2015} 
are conceptually closest to WEX, as both use 
regular expressions to select intervals over a sequence.
The key distinction 
is that spanners work in an offline setting; they assume access to the whole 
document. In contrast, our setting is online, and the window boundaries 
must be determined in a \textit{$0$-delay manner}, i.e., windows are selected based 
on data up to the current position and cannot depend on unseen future data. Another 
key difference in formalism is that WEX supports alphabet theories,
allowing window selection based on predicates over an infinite domain.

\section{Symbolic Automata, Expressions and MSO}
\label{sec:prelim}

We assume that data streams are infinite sequences of letters from an 
infinite alphabet $\Sigma$. 
Given  a word $w \in \Sigma^*$ and 
$i,j \in \mathbb{N}$, 
$w[i:j]$ represents the contiguous substring of $w$ starting from  
$i$\textsuperscript{th} index to $j$\textsuperscript{th} index, both 
inclusive. We start indexing from $0$. Let $w[i] = w[i:i]$ and $w[:k]$ be 
the suffix of $w$ of length $k+1$.

We introduce the model of $k$-symbolic automata, which is a variant of symbolic
automata with lookback. 
We refer  to \cite{SymbolicAutomata,SymbolicMSO,SymbolicRegEx} for details. In 
standard 
automata and MSO, transitions, and atomic formulas can check 
that the symbol at a position is equal to some particular letter in a finite 
alphabet. In Symbolic automata and MSO, we can instead check that the 
symbol at a position satisfies some property specified in 
the first-order 
logic. For example, 
an atomic formula of symbolic MSO can check that an input symbol is 
an even number, which can be specified in first-order logic over $\Nat$ 
with addition.

\begin{definition}
An \textbf{alphabet theory} is a tuple $\mathcal{A} = (\Sigma, V, 
\Psi_V)$ such that $\Sigma$ is a (finite or infinite) alphabet and $\Psi_V$ is a set 
of 
first-order formulas with free variables $V$ closed under boolean 
connectives $\{\lor, \land, \neg\}$ with $ \bot, \top \in \Psi_V$. 
Given $\psi \in \Psi_V$ and a valuation $\nu: V \to \Sigma$, it should 
be decidable to check whether $\nu \models_\mathcal{A} \psi$.
\end{definition}

We use $\models_\mathcal{A}$ to denote the models relation in the 
alphabet theory, to distinguish it from models relation in other logics. 
We use a variation of symbolic automata that can read symbols in the 
previous $k$ positions, for a fixed $k$, in addition to the current 
symbol. We call them $k$-symbolic lookback automata ($\ksla$), 
similar to $k$-symbolic lookback transducers introduced in 
\cite{d2015extended}.
\begin{definition}
	A $\mathbf{k}$-\textbf{symbolic lookback automata} ($\ksla$) is a
	tuple $S = (\mathcal{A}, Q, q_0, F, \delta)$ where 
	$\mathcal{A} = (\Sigma,V, \Psi_V)$ is an alphabet theory with $V = 
	(x_{-k},\ldots, x_0)$ as a set of $k+1$ 
	\emph{lookback variables}, $Q$ is a finite set of states, $q_0 \in Q$ 
	is the initial state, $F \subseteq Q$ is the set of final states, and 
	$\delta:  
	Q \times Q \to \Psi_V$ is the transition function.

	Given a word $w$ of length $|w| = k+1$, $\val_V(w): V \mapsto \Sigma$ denotes a 
	valuation such that $\val_V(w)(x_{-j}) = w[k -j]$ for all $j \in 
	[0,k]$.
	We define the \textbf{run} of $S$ on a word $w \in 
	\Sigma^*$ of length $n \ge k+1$ to be a sequence $(q_0, q_1, \ldots,  
	q_{n - k - 1})$ of states such that it starts from the initial state $q_0$, 
	and for all $i \in [k, n-1]$ we have $\val_V(w[i-k:i]) 
	\models_\mathcal{A} \delta(q_{i-k}, 
	q_{i-k+1})$.

	If $q_{n-k-1} \in F$, then the run is 
	accepting, and the word is \textbf{accepted}
	by $S$. If there is no accepting run, the word is \textbf{rejected}. 
	The run is not defined for words of 
	length less than 
	$k+1$, which are all rejected by a $\ksla$.

	The \textbf{language} accepted by $S$ is $\lang{S} = \{w 
	\in \Sigma^* \mid w \textit{ is accepted by } S\}$.	
\end{definition}

\begin{definition}
If $\delta(q_i, q_j) = \varphi$, then $\varphi$ is called the 
guard of the transition $q_i \xrightarrow[]{\varphi} q_j$. A $\ksla$ is 
\textbf{deterministic} if for every $q, q',q'' \in Q$ with $q' 
\ne q''$, $\delta(q,q') \land \delta(q,q'')$ is unsatisfiable. 

We say a $\ksla$ is \textbf{clean} if every $\delta(q,q')$ is either 
$\bot$ or is satisfiable. We can easily construct an equivalent clean 
$\ksla$ for any $\ksla$ by replacing all the unsatisfiable guards by 
$\bot$. 
\end{definition}

\para{$\mathbf{k}$-concatenation}
We define a variant of concatenation, 
called $k$-concatenation, denoted by $\cdot_k$. 

\begin{definition}
Given two strings $w_1 = 
wv$ and $w_2 = vw'$ 
and $|v| = k$, we 
define the \textbf{$\mathbf{k}$-concatenation} $w_1 \cdot_k w_2 = wvw'$. If the last $k$ letters of $w_1$ do 
not exactly match with first $k$ letters of $w_2$, or length of either 
word is less than $k$, then the concatenation is undefined. 
The $k$-concatenation of two languages $L_1$ and $L_2$ is defined as 
$L_1 \cdot_k L_2 = \{wvw' \in \Sigma^* \mid  wv \in L_1, vw' \in L_2  
\text{ and } |v|  = k\}$.
\end{definition}
\begin{example}
For $2$-concatenation, $ab\textcolor{green!20!black}{aa} \cdot_2 \textcolor{green!20!black}{aa} ba = ab\textcolor{green!20!black}{aa} ba$, while $ab\textcolor{red}{ab}  \cdot_2 \textcolor{green!20!black}{aa} ba$ is undefined.
\end{example}

The usual closure 
properties are satisfied by $\ksla$, which we establish next.
The results given in this section 
are easy adaptations of similar results for finite alphabets.

\begin{lemma}[Determinization]
	\label{lem:determinization}
Given a $\ksla$ $S  = (\mathcal{A}, Q, q_0, F, \delta)$, we can 
construct a deterministic $\ksla$ $S' = (\mathcal{A}, Q', q_0', F', 
\delta')$ such that  $\lang{S} = \lang{S'}$. 
\end{lemma}

\begin{proof}
We can assume, WLOG, that $S$ is clean. Let $q \in Q$ and $\textbf{q} 
\subseteq Q$. We will first define the following useful notation. 
\begin{align*}
    \delta_S(q) &= \{ (q, \delta(q,q'), q') \ne \bot \mid q' \in Q\}\\
    \delta_S(\textbf{q}) &= \cup_{q \in \textbf{q}} \delta_S(q)\\
    Target(\textbf{t}) &= \{ q' \mid (q, \varphi, q') \in \textbf{t}\}\\
    Cond((q, \varphi, q')) &= \varphi 
\end{align*}
Let $Q' = 2^Q$. To define the transition function $\delta'$, we will 
define the outgoing transition from each $\textbf{q} \subseteq Q$. For 
each subset $\textbf{t} \subseteq \delta_S(\textbf{q})$  let 
$$\varphi_\textbf{t} = (\bigwedge\limits_{t \in \textbf{t}} Cond(t)) 
\land (\bigwedge\limits_{t \in \delta_S(\textbf{q})\setminus \textbf{t}} 
\lnot Cond(t))$$. 

If $\varphi_\textbf{t}$ is satisfiable, then define $\delta'(\textbf{q}, 
Target(\textbf{t})) = \varphi_T$, otherwise $\delta'(\textbf{q}, 
Target(\textbf{t})) = \bot$. For all sets $\textbf{q'} \subseteq Q$ which 
are not equal to $Target(\textbf{t})$ for any $\textbf{t} \subseteq 
\delta_S(\textbf{q)}$, we define $\delta'(\textbf{q}, \textbf{q'})) = \bot$.

Define $q_0' = \{q_0\}$ and $F' = \{\textbf{q} \subseteq Q \mid 
\exists q \in \textbf{q}. q \in F\}$. 

To see that $S'$ is deterministic, note that for every two outgoing 
transition from a state $\textbf{q} \in Q'$ with guard 
$\varphi_\textbf{t}$ and $\varphi_\textbf{t'}$ with $\textbf{t} \ne 
\textit{t'}$, $\varphi_\textbf{t} \land \varphi_\textbf{t'}$ is 
unsatisfiable. It follows because, WLOG, if $t \in \textbf{t}$ and $t 
\notin \textbf{t'}$, then $\varphi_\textbf{t}$ will have the conjunct 
$Cond(t)$ while $\varphi_\textbf{t'}$ will have $\lnot Cond(t)$. 

It's trivial to verify that $S'$ accepts the same language as $S$.
\qed
\end{proof}

\begin{definition}
We complement languages of $\ksla$s with respect to strings of length 
at least $k+1$. The \textbf{complement} of a language of $\ksla$ is defined as 
$\overline{\lang{S}}= (\Sigma^{k+1} \cdot\Sigma^*) 
\setminus \lang{S}$. The following result follows from the previous one.
\end{definition}

\begin{lemma}[Complementation]
	\label{lem:complementation}
Given a $\ksla$ $S$, we can construct a $\ksla$ $S'$ such that 
$\lang{S'} = \overline{\lang{S}}$.
\end{lemma}
\begin{proof}[Sketch]
	Proof is similar to the standard NFA complementation procedure. 
	We first determinize $S$ and then swap the final states
	with non-final states. 
	\qed
\end{proof}

Product construction works on $\ksla$ as usual. Given two $\ksla$ $S_1 
= (\mathcal{A}, Q_1, q_{0}^1, 
F_1, \delta_1)$ and $S_2 = (\mathcal{A}, Q_2, q_0^2, F_2, \delta_2)$, 
the product of $S_1$ and $S_2$ is $S = (\mathcal{A}, Q_1 \times Q_2, (q_{0}^1, 
q_0^2), F, \delta)$ where $F = \{(q_1, q_2) \mid q_1 \in F_1 \textit{ and } q_2 \in 
F_2\}$ and $\delta((q_1, q_2), (q_1', q_2')) = \delta_1(q_1, q_1') \land 
\delta(q_2, q_2')$ for all $q_1, q_1' \in Q$ and $q_2, q_2' \in Q_2$.

\begin{lemma}[Intersection]
Given two $\ksla$ $S_1 = (\mathcal{A}, Q_1, q_{0}^1, F_1, \delta_1)$ 
and $S_2 = (\mathcal{A}, Q_2, q_0^2, F_2, \delta_2)$, let the product 
of $S_1$ and $S_2$ be $S$. Then, $\lang{S}
= \lang{S_1} \cap \lang{S_2}$. 
\end{lemma}

The languages of $\ksla$ are also closed under union, which can be 
proved as usual by taking disjoint union of two automata.

\begin{lemma}
	\label{lem:concatenation}
The languages of $\ksla$ are closed under $k$-concatenation.
\end{lemma}

\begin{proof}
	Let $\ksla$ $S_1 = (\mathcal{A}, Q_1, q_{0}^1, F_1, \delta_1)$ and 
	$S_2 = (\mathcal{A}, Q_2, q_0^2, F_2, \delta_2)$. The construction 
	is similar to that in case of finite automata. Formally, construct a 
	$\ksla$ $S = (\mathcal{A}, Q_1 \cup Q_2, q_0^1, F_2, \delta)$ with
	$$
	\delta(q, q') = 
	\begin{cases}
	\delta_1(q,q'), &\text{if } q,q' \in Q_1\\
	\delta_2(q,q'), &\text{if } q,q' \in Q_2\\
	\delta_2(q_0^2,q'), &\text{if } q \in F_1, q' \in Q_2 \\
	\bot, &\text{otherwise}
	\end{cases}
	$$
	In the third case above, $S$ non-deterministically switches from 
	$S_1$ 
	to $S_2$. The proof of correctness is routine.
	\qed
\end{proof}

\subsection{Symbolic Regular Expressions}

\para{Symbolic Regular Expressions (SRE)} SREs are regular expressions over an
alphabet theory. 

\begin{definition}
Given an alphabet theory $\mathcal{A} = (\Sigma \cup 
\{\emptylookback\}, V, \Psi_V)$, the set of \textbf{Symbolic Regular 
Expressions} is defined by the  grammar: $$R \coloneq [\varphi]  
~|~ R + R ~|~ R \cdot_k R ~|~ R^*,$$ 
where $\varphi \in \Psi_V$.
\end{definition}
The semantics of SRE are defined as follows:
$\sem{[\varphi]} \coloneq \{w \in \Sigma^{k+1} \mid
\val_V(w) \models_\mathcal{A} \varphi\}$, $\sem{R_1 + R_2} \coloneq \sem{R_1} \cup \sem{R_2}$, 
$\sem{R_1 \cdot_k R_2} \coloneq \sem{R_1} 
\cdot_k \sem{R_2}$ and $\sem{R^*} \coloneq 
\bigcup\limits_{1 \le n \in 
	\mathbb{N}} \sem{R^n}$.

\begin{lemma}
	For every SRE $R$ there is a $\ksla$ S (and vice versa) such that $L(S) = L(R)$.
\end{lemma}
\begin{proof}[omitted] Trivial proof via same constructions as for finite alphabet.
\end{proof}

\subsection{S-MSO: Symbolic Monadic Second Order}

We define Symbolic Monadic Second Order (S-MSO) logic from 
\cite{SymbolicMSO} with support for lookback variables.
\begin{definition}
	Given an alphabet theory $\mathcal{A} = (\Sigma, 
	V, \Psi_V)$, The syntax of S-MSO 
	over $\mathcal{A}$ is defined by the following grammar:
	$
	\phi ::= [\varphi](x) ~|~ 
	x < y ~|~
	X(x) ~|~
	\lnot \phi ~|~
	\phi \land \phi ~|~ 
	\exists x ~\phi ~|~
	\exists X~ \phi
	$, where $\varphi \in \Psi_V$, lower case letters $x,y,z$ are 
	first-order 
	variables and upper case letters $X,Y,Z$ are second order variables. Note that the lookback variables in $V$ can 
	occur only in formula from $\Psi_V$.
\end{definition}

\para{Semantics of S-MSO}
Let $\phi$ be an S-MSO formula with free variables $FV(\phi)$. 
Consider a word  $w 
\in \Sigma^*$ with $|w| \ge k+1$ and a map $\theta: 
FV(\phi) \to [k, |w|-1] \cup 2^{[k, |w|-1]}$, where the first order 
variables are mapped to $[k, |w|-1]$ and second order variables are 
mapped to $2^{[k, |w|-1]}$. Given a substring $w' = a_0\ldots a_k$ of $w$ of size $k+1$, 
we define $\nu[w']: V \to \Sigma$ to be a map with $\nu[w'](x_{i-k}) = a_i$ for 
all $i \in [0,k]$. 
The semantics of S-MSO with 
$k$-lookback is as follows.

\begin{alignat*}{2}
&w, \theta \models [\varphi](x) &&\Leftrightarrow \val_V(w[\theta(x) 
- k: \theta(x)]) \models_\mathcal{A} \varphi\\
&w, \theta \models x < y &&\Leftrightarrow \theta(x) < \theta(y)\\
&w, \theta \models X(x) &&\Leftrightarrow \theta(x) \in \theta(X)\\
&w, \theta \models \lnot \phi &&\Leftrightarrow w , \theta 
\not\models \phi\\
&w, \theta \models \phi_1 \land \phi_2 &&\Leftrightarrow w, \theta 
\models \phi_1 \textit{ and } w , \theta \models \phi_2\\
&w , \theta \models \exists x ~\phi(x) &&\Leftrightarrow \exists i 
\in 
[k, |w|-1] \textit{ such that } w , \theta[x\mapsto i] \models \phi(x)\\
&w , \theta \models \exists X ~\phi(X) &&\Leftrightarrow \exists I 
\in 
2^{[k, |w|-1]} \textit{ such that } w , \theta[X \mapsto I] \models 
\phi(X)
\end{alignat*}

\para{Equivalence of S-MSO and $\ksla$ 
}\label{sec:equiv-wmso-ksla}
To prove that $\ksla$ and S-MSO are equally expressive, the 
following extension of alphabet theories is helpful.
\begin{definition}
	The \textit{extension} of an alphabet theory $\mathcal{A} = (\Sigma, 
	V, \Psi_V)$ with a boolean variable $x$ is 
	a new alphabet theory $\mathcal{A}_{x} = (\Sigma \times \{0,1\} , V, 
	\Psi_V \times \{x=0, x=1\})$ such that for any $a 
      	\in \Sigma$ and $b,b' \in \{0,1\}$, $(a,b) \models (\varphi, x=b')$ if 
	and only if $a \models_{\mathcal{A}} \varphi$ and $b = b'$.
	We denote the extension by $n$ boolean variables $x_1, x_2, \ldots, 
	x_n$ by $\mathcal{A}_{(x_1,x_2, \ldots, x_n)}$.
\end{definition}

Let $x_1, \ldots, x_n$ denote first order variables and $X_1, \ldots, X_m$
denote the second order variables for an S-MSO formula $\phi$. 
Given a word $w \in \Sigma^*$  and a map $\theta $, let us define a 
word over extended alphabet theory $w_\theta \in 
(\Sigma \times 
\{0,1\}^{n+m})^*$ as 
$$w_\theta[i] = (w[i], e_1(\theta(x_1), i), \ldots, 
e_1(\theta(x_n), i), e_2(\theta(X_1), i), \ldots e_2(\theta(X_m), i)),$$ 
where $e_1(n, i) = 1$ if $n = i$ else $0$, and $e_2(I, i)= 1$ if $i \in I$ 
else $0$. Recall that this is similar to the usual construction one does 
while showing equivalence of MSO and finite automata. 

\begin{lemma}
	Let  $\mathcal{A} = (\Sigma, V, \Psi_V)$ be an alphabet theory, and $w \in \Sigma^*$ be any word. 
	\begin{itemize}
		\item For every S-MSO formula $\phi$ over the free variables $\{x_1, \ldots, x_n, X_1, \ldots, X_m\}$,
		there exists an $\ksla$ S over the extended alphabet theory $\mathcal{A}_{(x_1, \ldots, x_n, X_1, \ldots, X_m)}$
		such that $w \models \phi \iff w \in L(S)$.
		\item For every $\ksla$ $S$ there exists an equivalent S-MSO sentence $\phi$ such that
		$w \in L(S) \iff w \models \phi$.
	\end{itemize}
	
\end{lemma}

\begin{proof}[sketch] 
Let $\phi(x_1, \ldots, x_n, X_1, \ldots, X_m)$ be a S-MSO formula with 
first order free variables $\{x_1, \ldots, x_n\}$ and second order free 
variables $\{X_1, \ldots, X_n\}$.
A construction similar to that of MSO over finite alphabets will give a 
$\ksla$ $S$ over the extended alphabet theory $\mathcal{A}_{(x_1, 
	\ldots, x_n, X_1, \ldots, X_m)}$, such that $w,\theta 
\models \phi$ if and only if the word $w_\theta \in (\Sigma \times 
\{0,1\}^{n+m})^*$ is 
accepted by $S$.

Conversely, given a $\ksla$ over the alphabet theory $\mathcal{A} = 
(\Sigma, V, \Psi_V)$, 
we can construct a S-MSO formula $\phi$ over $\mathcal{A}$ with no 
free variables such that $w \in \Sigma^*$ is accepted by $S$ if and 
only if $w \models \phi$ using the standard automata to MSO 
construction and replacing the letters with predicates $\varphi \in 
\Phi_V$.
\qed
\end{proof}

The expressive power of symbolic lookback automata and symbolic 
MSO are useful for designing parsing algorithms and specification 
languages as we will see subsequently. But, unfortunately, 
the expressive power is 
enough to simulate Turing machines and static analysis problems are 
undecidable.
\begin{theorem}
	The problems of checking non-emptiness of languages of $\ksla$s 
	and satisfiability of S-MSO formulas are undecidable.
\end{theorem}
\begin{proof}[sketch]
The $\ksla$ automata work over infinite domains and transitions can relate 
values at a position with previous values. This can be used to 
simulate counter machines. The domain is the set $\mathbb{N}$ of 
natural numbers and the counters are simulated by 
numerical fields in the input stream. An incrementing  transition of 
the counter 
machine can be simulated by a transition 
of a $\ksla$, by requiring that the next value of 
the corresponding field is one more than the previous one. 
Decrementing and zero testing transitions can be similarly simulated. 
This is a standard trick used for models dealing with infinite 
domains, e.g., \cite{d2015extended,AutomataApproachCLTL}.	
\end{proof}

\section{WEX: Defining Windows with S-MSO}
\label{sec:WindowsSMSO}
A window in a data stream is a pair $(i_b,i_e)$ of indices that indicate 
where the window begins and ends. 
In this section, we explain how windows can be defined with S-MSO 
over an alphabet theory $\mathcal{A}$, and also show an expressively 
equivalent representation using symbolic regular expressions. 

We designate first-order variables $x_b,x_e$ for denoting the 
beginning and ending indices of windows. We use S-MSO formulas to 
specify which indices can begin and end windows. The end of a window 
should be detected as soon as it arrives in the stream, so the decision 
about whether a position is the end of a window should be made based 
only on the stream data that has been read so far. We enforce this in 
S-MSO formulas by guarding the quantifiers.
	\begin{align}
		\phi &:= [\varphi](x) ~|~ 
		x < x' ~|~
		X(x) ~|~
		\lnot \phi ~|~
		\phi \lor \phi
		~|~ \exists x \le x_e  ~\phi ~|~
		\exists X \subseteq [0,x_e]  ~ \phi	
		\label{eq:WSMSO}
	\end{align}
The above syntax is a guarded fragment of S-MSO --- $\exists x \le 
x_e  ~ \phi$ is syntactic sugar for $\exists x ( x\le x_e  ~ \land ~\phi)$ 
and $\exists X \subseteq [0,x_e]  ~ \phi$ is syntactic sugar for $\exists 
X ( \forall y (X(y) \Rightarrow y \le x_e) ~ \land ~ \phi)$.

\begin{definition}
Let $x_b \le x_e ~ \land ~\phi(x_b,x_e)$ be a S-MSO formula in the 
guarded fragment given in \eqref{eq:WSMSO}, with $x_b$ and $x_e$ 
being free 
variables. A pair $(i_b,i_e)$ of indices of a word $w$ is said to be a 
window recognized by $x_b \le x_e ~ \land ~\phi(x_b,x_e)$ if $i_b \le 
i_e$ and $w \models \phi(i_b,i_e)$

\end{definition}
To reduce clutter, we don't explicitly write the condition $x_b 
\le x_e$ but assume that it is present in all window 
specifications.
Whether a pair $(i_b, i_e)$ is recognized as a window by $\phi(x,y)$ 
in a word $w$ depends only on the word $w[0:i_e]$.

End users of streaming data processors may not be familiar with logic 
based 
languages. It's a general observation that specifications 
based on regular expressions are easier to understand compared to 
those based on logic. Next we give a way of defining windows based 
on symbolic regular expressions, that is expressively equivalent to the 
one above based on S-MSO.

\begin{definition}
A Window EXpression (\textbf{WEX}) is a set 
$\mathcal{R} = \{(r_1, r_1'), \ldots (r_l, r_l')\}$. For every $i$, $r_i, r_i'$ 
are 
symbolic regular expressions over an alphabet theory $\mathcal{A}$.

Given a word $w \in \Sigma^*$, a pair $(i_b,i_e) \in [k, |w|-1]^2$ is 
said to be \textbf{recognized as a window} by $\mathcal{R}$ if there exists $(r, 
r') \in \mathcal{R}$ such that $w[0:i_b-1] \in \sem{r}$ 
and 
$w[i_b-k : i_e] \in \sem{r'}$.
\end{definition}

Intuitively, one can think of $r$ as a \textit{prefix
automaton} that decides where the window can begin,
and $r'$ as a \textit{window automaton} that decides
where the window can end. We
need to include the $k$ proceeding symbols before $i_b$ 
as $r'$ is a $\ksla$ that needs access to 
$k$ symbols to start reading from index $i_b$.

\begin{lemma}
Given a S-MSO formula $\phi(x_b,x_e)$ in the guarded fragment given 
in \eqref{eq:WSMSO}, we can effectively construct a window expression 
$\mathcal{R} = 
\{(r_1, r_1'), \ldots ,(r_l, r_l')\}$
such that for every word $w$, the set of windows recognized by 
$\phi(x_b,x_e)$ is the same as that recognized by $\mathcal{R}$.
\label{lem:WSMSOToExpr}
\end{lemma}

\begin{proof}
The idea is to take the automaton corresponding to the S-MSO formula $\phi$ 
and split it at a transition that reads the symbol at position $x_b$. 
Each such split results in a pair of expressions.

From section \ref{sec:prelim}, given an S-MSO formula $\phi(x_b,x_e)$, 
we can construct a $\ksla$ $S$ over the extended alphabet theory 
$\mathcal{A}_{(x_b,x_e)}$, such that for any word $w \in 
\Sigma^k\Sigma^*$ and any map $\theta: \{x_b,x_e\} \to [k, |w|-1]$, 
$w,\theta 
\models \phi$ if and only if the word $w_\theta \in (\Sigma \times 
\{0,1\}^{2})^*$ is 
accepted by $S$.

Guards of transitions in $S$ are of the form 
$(\varphi,x_b=\alpha_b,x_e=\alpha_e)$, where $\varphi$ is a formula 
from the alphabet theory and $\alpha_b,\alpha_e \in \set{0,1}$. Let us 
denote by $S\downarrow x_b = 0$ the $\ksla$ obtained from $S$ 
by removing transitions whose guards have $x_b=1$. Similarly, we 
define $S\downarrow x_e = 0$.
For every pair of transitions $(t_b, t_e)$ in $S$ such that the guard of 
$t_b$ 
(resp.~$t_e$) has $x_b=1$ (resp.~$x_e=1$), we construct the 
following two $\ksla$s:
\begin{enumerate}
	\item $\pa_{(b,e)}$: We start with the $\ksla$ $S\downarrow 
	x_b=0$. Set 
	the initial state to be the same as that of $S$. Add a new state $p_f$ 
	and set it as the only final state. Suppose the transition $t_b$ is from 
	state $q$ to $p$. In $\pa_{(b,e)}$, add a transition from $q$ to 
	$p_f$ with the same guard as $t_b$. This is intended to accept 
	prefixes of windows.
	\item  $\wa_{(b,e)}$: We start with the $\ksla$ $S\downarrow 
	x_e=0$. Set 
	$p$ as the initial state, where $p$ is the target state of $t_b$. Add a 
	new state $p'_f$ and set it as the only final state. Suppose the 
	transition $t_e$ is from $q'$ to $p'$. In $\wa_{(b,e)}$, add a 
	transition from $q'$ to $p'_f$ with the same guard as $t_e$. This is 
	intended to accept windows.
\end{enumerate}

We denote by $\sre(S)$ the symbolic regular expression equivalent to 
the $\ksla$ $S$. Define $\mathcal{R} = \{(SRE(\pa_{(b,e)}), 
SRE(\wa_{(b,e)})) \mid \text{guard of } t_b \text{ has }x_b=1, \text{ 
	guard of } t_e \text{ has }x_e=1\}$. We shall now prove that for all 
words $w \in 
\Sigma^k\Sigma^*$, the set of windows recognized in $w$ by 
$\phi(x,y)$ and $\mathcal{R}$ are same.

$\Rightarrow$: Let $(i_b, i_e)$ be any window recognized by 
$\phi(x_b,x_e)$ in $w$, so $w[0:i_e], \{x_b \mapsto i_b	, x_e 
\mapsto i_e\} 
\models \phi(x_b,x_e)$. Let $w' = w[0:i_e]$. By construction 
$w'_\theta$ will be accepted by the $\ksla$ $S$. Consider an accepting 
run $\rho := q_0 \trans{(*,0,0)} q_1 \ldots q_{i_b} \trans{(*, 1, *)} 
q_{i_b+1} \ldots q_{i_e-1} \trans{(*,*,1)} q_{i_e}$ of $S$ on $w'_\theta$.
As $x_b,x_e$ are first-order variables, there would be exactly two 
positions, $i_b$ and $i_e$ where they take the value $1$ respectively 
in 
the word $w_\theta$, and elsewhere they would be $0$.

Consider the pair $(SRE(\pa_{(b,e)}),SRE(\wa_{(b,e)})) \in \mathcal{R}$ 
where $t_b = q_{i_b} \trans{(*,1,*)} q_{i_b+1}$ and $t' = q_{i_e-1} 
\trans{(*,*,1)} q_{i_e}$. It follows by construction that $\pa_{(b,e)}$ 
accepts $w[0:i_b-1]$ and $\wa_{(b,e)}$ accepts $w[i_b -k: i_e]$. 

$\Leftarrow$: Let $(i_b, i_e)$ be a window accepted by the pair 
$$(SRE(\pa_{(b,e)}), SRE(\wa_{(b,e)})) \in \mathcal{R}$$ with  $t_b := q 
\trans{(*,1,0)} p$ and $t_e = q' \trans{(*,0,1)} p'$. Therefore, we have 
$w[0:i_b-1] \in L(\pa_{(b,e)})$ and $w[i_b -k: i_e] 
\in L(\wa_{(b,e)})$. Let $\rho_1 = q_0 \trans{(*,0,0)} q_1 \ldots 
q_{i_b-1} 
\trans{(*, 1, *)} q_{i_b} = p_f $ and $\rho_2 =  q_{i_b} \trans{} 
q_{i_b+ 1} \ldots \trans{(*,*,1)} q_{i_e} = p'_f$ be the accepting runs of 
the $\ksla$s 
$\pa_{(b,e)}, \wa_{(b,e)}$ on $w[0:i_b-1], w[i_b -k: i_e]$ respectively. 
We can combine the runs 
$\rho_1$ and $\rho_2$ by replacing the last transition in the runs with 
$t_b,t_e$ respectively and merging the runs to get a run $\rho$ in $S$ 
for 
the word $w[0:i_e]_\theta$ with $\theta := \{x \to i_b, y \to i_e\}$.
\end{proof}

\begin{lemma}
Given a WEX $\mathcal{R} = \{(r_1, r_1'), \ldots, (r_l, 
r_l')\}$ over $\mathcal{A}$, we can effectively construct a S-MSO 
formula 
$\phi(x_b,x_e)$ in the guarded fragment given in \eqref{eq:WSMSO} 
such that, for 
every word $w$, the set of windows recognized by $\mathcal{R}$ is 
same as that recognized by $\phi(x_b,x_e)$.
\label{lem:ExprToWSMSO}
\end{lemma}

\begin{proof}
	Consider a pair $(r, r') \in \mathcal{R}$. Let $\psi, \psi'$ be the 
	S-MSO 
	sentences corresponding to $r,r'$ respectively. The sentence $\psi$ 
	partitions the set of positions of a word into multiple parts, each part 
	corresponding to a state of the automaton for the expression $r$. 
	The 
	sentence $\psi$ further verifies that the partition forms a valid run 
	according to the transition rules of the automaton. We modify $\psi$ 
	to 
	partition the set of postions $[0,x_b-1]$ instead and verify the 
	validity 
	of the partition. Let us call this modified formula 
	$\psi\upharpoonright 
	x_b$. It is routine to verify that $\psi\upharpoonright x_b$ can be 
	written in the guarded fragment described above. Similarly, 
	$\psi'\upharpoonright[x_b:x_e]$ will partition and verify the 
	positions 
	in $[x_b,x_e]$. The required formula $\phi(x_b,x_e)$ for $(r,r')$ is 
	$(x_b \le x_e) 
	\land (\psi\upharpoonright x_b) \land (\psi'\upharpoonright 
	[x_b:x_e])$. The required formula is the disjunction of all such 
	formulas 
	for 
	all 
	the pairs in $\mathcal{R}$.
	\qed
\end{proof}

\subsection{Examples from Practical Applications}\label{sec:examples}
In this section, we give a few examples from practical applications of 
stream processing, to illustrate how WEX can express queries with 
clearly specified semantics.
Let us denote the atomic formula in alphabet theory that is
always true by $\top$ with $L(\top) = \Sigma^{k+1}$.

\para{1. Trends in stock markets} 
Continuing the motivating example from the introduction,
we formalize the windowing construct with WEX by fixing the theory of 
reals with lookback variable set $V = \{x_{-2}, x_{-1}, x_{0}\}$ as the alphabet  
theory. 
We first define the following atomic formulas: $D_2 := x_{-2} > x_{-1} \land x_{-1} > x_{0}$
and 
$U_1 := x_{0} > x_{-1}.$
The formula $D_2$ is true at a position if it follows two consecutive decrements in the 
average stock price, and $U_1$ is true at a position if the stock price has just increased.
The WEX $R = (\pa, \wa)$ where $\pa = 
 \top^*\cdot_3 [D_2]$ and $\wa = U_1\cdot_3[\lnot D_2]^* \cdot_3 [D_2]$ precisely
 specifies the windowing construct
illustrated in~\Cref{figure:stock-trend-example} that we wished 
 to capture.

\para{2. Arrhythmia Detection} 
An ICD delivers shocks to restore normal heartbeat when 
it detects arrhythmia \cite{ARMBSG2019}.  
For our purposes, the input stream is a sequence
 $\{h_i\}_i$, where $h_i$ is the $i$th measurement from a 
 heartbeat spectrogram. A sequence value $h_i$ is \textit{peaking} if
  $h_i \ge \max(h_{i-1},h_{i+1})$ and $h_i \ge p$, for some
predefined threshold $p$.  

We fix the theory of reals
with $k=2$ and three lookback variables $V = \{x_{-2}, x_{-1}, x_0\}$ as the alphabet theory. 
Let
$R = (\pa, \wa)$ be a WEX, where $\pa 
= \top^*$ and 
$\wa = [x_{-1} > x_{-2} \land 
x_{-1} > x_0 \land 
x_{-1} > p]$. 
On processing an input stream with $R$, the output will be a stream of windows
$w_i = hh'$, such that the first value $h$ is 
peaking.

\section{Window Overlap Problem}\label{section:window-overlap-problem}
Consider the WEX 
$((a+b)^*,a^*b)$ over the finite alphabet $\Sigma=\set{a,b}$. A window can start at any position but must end at the 
letter 
$b$. 
Applied to the input stream $a^nb$, this WEX accepts $n+1$ windows, each of the form $a^ib$ for $i \in [0,n]$,
all mutually overlapping.
However, if we restrict the input streams to those that do not have $a^c$
as an infix, for some constant $c$,
the same WEX will always have a bound on the number
overlapping windows. In particular, there are at most $c$ overlapping windows at any time.

A natural problem arises to determine whether a given WEX allows
unbounded number of overlapping windows.
The answer to this depends on whether the input data stream follows any
pattern in relation to the window expressions.
This motivates formally defining the class of streams under consideration.
As we will see later, other than the theoretical nicety,
this problem has a practical importance in determining whether a stream processor 
can require unbounded amount of memory when processing a stream for a given WEX.

\para{Input Specifier for Streams}
To formalize which streams are to be considered, we use \textbf{input specifiers}, 
which are prefix-closed $\ksla$s, i.e., if $w$ is accepted then every prefix of $w$
of length greater than $k$ is also accepted. We can think of a specifying automaton $S$ as 
being in an ``accepting zone'' or in a ``rejecting zone'', it starts in an
accepting zone and switches to the rejecting zone at most once. A stream is said to \textit{conform} to an input 
specifier if it never enters the rejecting zone while reading the stream.

\begin{definition}[Window Overlap Problem] 
    Two windows \textbf{overlap} if the intervals defining them intersect.
    Given a WEX $R$ and an input specifier $S$ over alphabet theory $\mathcal{A}$, determine whether for each $n \in \mathbb{N}$,
    there exists a stream $s_n$ conforming to $S$ such that there exists at least $n$ mutually overlapping
    windows accepted by $R$ over the stream $s_n$.
\end{definition}

In general, the window overlap problem is undecidable, just like the 
problem of checking non-emptiness of languages of $\ksla$s. We 
identify a fragment for which it is decidable. 

\begin{definition}
An alphabet 
theory $\mathcal{A} = (\Sigma,V, \Psi_V)$ is called \textbf{simple} 
if $\Psi_V$ is restricted to boolean combination of atomic formulas 
of the form $R(x_1,\ldots, x_l)$, where $R$ is a relation symbol. In particular, 
we don't allow quantifiers in formulae.
\end{definition}

Checking non-emptiness of the languages of a $\ksla$ over 
a simple alphabet theory is also
undecidable, if the equality relation and one more relation symbol are 
present (see \cite[Theorem 10.1]{AutomataApproachCLTL}). Hence, 
the window overlap problem also remains undecidable over them.

We further restrict simple alphabet theories by adapting the concept of 
the \textit{completion property} \cite[Section 4]{AutomataApproachCLTL}.

\begin{definition}
Suppose $\Phi$ is a set of atomic formulas over the set of variables 
$V$, $V' \subseteq V$ and 
$\Phi \restr V' \subseteq \Phi$ is the set of 
those formulas in $\Phi$  that only use variables in $V'$. 

A simple alphabet 
theory is said to have the \textbf{completion property} if for every satisfiable set 
of formulas $\Phi$, for every subset $V' \subseteq V$ and every partial 
valuation $\val':V' \to \Sigma$ that satisfies all the constraints in $\Phi 
\restr V'$, there exists an extension $\val: V \to \Sigma$ of $\val'$ that 
satisfies all the constraints in $\Phi$.
\end{definition}

The theory of integers with the 
binary relation $<$ does not satisfy the completion property. Consider 
the set of formulas $\set{x<y,x<z, z<y}$ and a partial valuation $\val': 
\set{x \mapsto 1, y \mapsto 2}$. It satisfies $x < y$, but it cannot be 
extended to include a mapping for $z$ such that $x < z$ and $z < y$ 
(i.e., $z$ is strictly between $x$ and $y$), since there is no integer 
strictly between $1$ and $2$. The theory of rational or real numbers 
with $<$ satisfy the completion property. For theories with linear 
orders, the completion property is closely related to the denseness of the 
domain \cite[Lemma~5.3]{AutomataApproachCLTL}. Note that the completion
property depends upon both the domain and the relations.
For example, theory of rational or real numbers with the binary relations
$<$ and $+$ (addition) does \textbf{not} have completion property.

\begin{theorem}\label{thm:wop-decidable}
    The Window Overlap Problem is decidable for WEX over simple alphabet 
    theory with completion property.\footnote{Recall that for finite alphabets, the set of formula $\Phi$ is an empty set.
Therefore, they trivially are
simple and have completion property.}
\end{theorem}

\begin{lemma}\label{lem:wop-characterization}
 	Given a WEX as a 
 	pair $(\pa, \wa)$ of 
 	deterministic $\ksla$s over an alphabet theory with the completion property,
    and an input specifier $S$, the number of 
    overlapping windows will be unbounded 
 	over streams in $S$ iff 
 	there exist words $w_1,w_2,w_3$ and a state $q$ of $\waß$ such that
 	\begin{enumerate}
 		\item There is a path from $q$ to a final state in which 
 		no transition has the guard $\bot$ (we say that $q$ is not a dead 
 		state in this case),
 		\item in the input specifier $S$, $\initS \trans {w_1} s \trans{w_2} 
 		s \trans{w_3}s$, such that $s$ is an accepting state,
 		\item $\initpa \trans{w_1} p\trans{w_2} p \trans{w_3} p$, where 
 		$p$ 
 		is a final state of $\pa$,
 		\item $\initwa \trans{w_1[:k]w_2} q$, $q \trans{w_1[:k]w_2} q 
 		\trans{w_3}q$ and
 		\item the runs $s \trans{w_1[:k]w_2} 
 		s \trans{w_3}s, p\trans{w_1[:k]w_2} p \trans{w_3} p, q 
 		\trans{w_1[:k]w_2} q 
 		\trans{w_3}q$ satisfy the following: the sequence of 
 		transitions used while reading the first $k$ letters of $w_2$ is 
 		the same sequence used for reading the first $k$ letters of $w_3$.
 	\end{enumerate} 
\end{lemma}
 \begin{proof}
 	($\Rightarrow$) 
 	Let $w$ be a stream conforming with $S$ for which there are 
 	at least 
 	$n$ overlapping windows. There must exist an 
 	increasing sequence of positions (or time instants) $(t_i)_i$ 
	corresponding to the start index of the overlapping windows,
    such that 
	$w_n[0:t_i] \in L(\pa)$.
    Let $w_n = w[0:t_n]$. 
 	
	\begin{figure}
		\centering
		\includegraphics[width=0.7\textwidth]{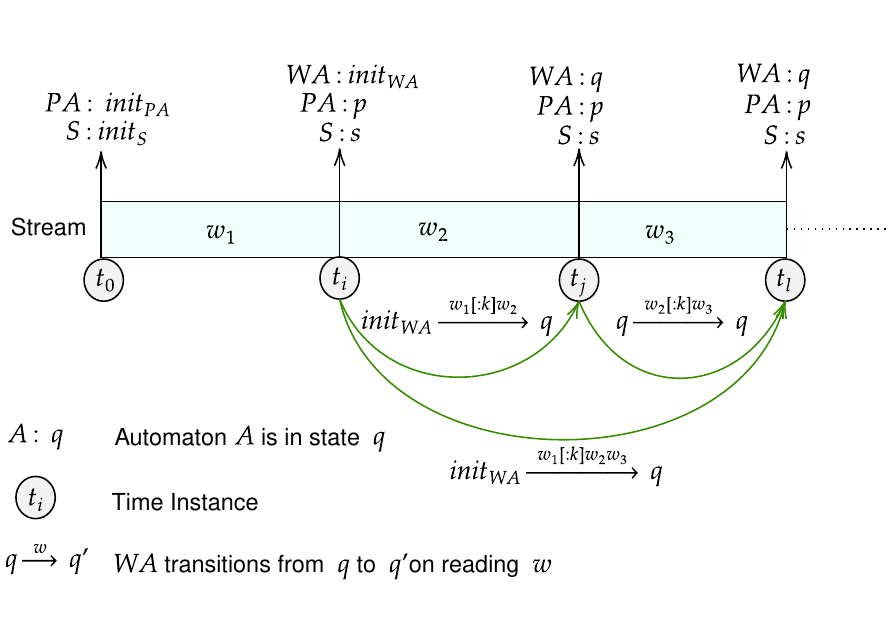}
		\caption{Illustration of the stream used in Proof of~\Cref{lem:wop-characterization}}\label{fig:illustration-stream-main-lemma}
	\end{figure}

 	Let $Q_\pa,Q_\wa, Q_S$ be the set of states in $\pa,\wa, S$ 
 	respectively. Let $T_\pa, T_\wa$ and  $T_S$ be the set of 
 	$k$-tuples of 
 	transitions in $\pa,\wa$ and $S$ respectively. Let $\mathbb{T}$ be 
 	the set 
 	of all functions of the form $Q_\wa \to T_\wa$. We will use the set 
 	of colors 
 	$C = \mathbb{T} \times 2^{Q_\wa \times 
 	Q_\wa}$ to color the edges of a graph we define later. Let $r$ be the 
 	Ramsey number $R(n_1, n_2, \ldots, n_{|C|})$, where $n_1 = n_2 = 
 	\ldots = n_{|C|} = 2|Q_\pa||Q_S||T_\pa||T_S| + 1$. We set $n = r$ 
 	and 
 	construct a complete graph $G$ with the set $\{t_i\}$ as vertices. For 
 	every two instants $t_i, t_j$ with $i < j$, we add an edge with color 
 	$(T_i,\{(q_1, q_2) \mid q_1, q_2 \in Q_\wa,  q_1 \trans{w_n[t_i+1-k: 
 	t_j]} 
 	q_2\})$, where $T_i \in \mathbb{T}$ is the function such that 
 	$T_i(q)$ is the sequence of $k$ transitions 
 	executed in $\wa$ if it starts at the instant $t_i$ in state $q$. We 
 	infer from Ramsey's 
 	theorem that the graph $G$ will have 
 	a monochromatic clique of size $2|Q_{PA}||Q_S||T_\pa||T_S| + 1$.

 	We infer from the pigeonhole principle that this monochromatic clique contains at least three time 
 	instants, say $t_i < t_j < t_l$, at which 
	$\pa$ and $S$ are in the same pair of states, say $p$ in $\pa$ and $s$ in $S$, both of which are accepting
	states in their respective automata.
	Also, the sequence of $k$ transitions executed in $\pa$ 
 	(resp.~$S$) from $t_i,t_j,t_l$ are same. 
	Now, consider 
 	the instances of window automaton $\wa$ initiated at the instants 
 	$t_i$, $t_j$ and $t_l$.
	
	~\Cref{fig:illustration-stream-main-lemma}
	may help parse the following argument.
 	Let $\initwa \trans{w_n[t_i+1-k:t_j]} q$ for some state $q$. Since the 
 	edges between $t_i,t_j,t_l$ all have the same color, $\initwa 
 	\trans{w_n[t_i+1-k:t_l]} q$. Since $\wa$ is deterministic and $t_i < 
 	t_j < t_l$, we can split $\initwa 
 	\trans{w_n[t_i+1-k:t_l]} q$ into  $\initwa \trans{w_n[t_i+1-k:t_j]} q$ 
 	and $q \trans{w_n[t_j+1-k: t_l]} q$. Since $(t_i,t_j)$ and $(t_j,t_l)$ 
 	have the same colour, we infer that
 	$q \trans{w_n[t_i+1-k: t_j]} q$. Let $w_1 = w_n[0:t_i]$, 
 	$w_2 = w_n[t_i+1:t_j]$ and $w_3=w_n[t_j+1:t_l]$. We have $\initpa 
 	\trans{w_1} p \trans{w_2} 
 	p \trans{w_3}p$, $\initS 
 	\trans{w_1} s \trans{w_2} 
 	s \trans{w_3}s$, $p$ (resp.~$s$) is accepting in $\pa$ (resp.~$S$), 
 	$\initwa \trans{w_1[:k]w_2} q$, $q \trans{w_1[:k]w_2} q \trans{w_3} 
 	q$ 
 	and the runs $s \trans{w_1[:k]w_2} 
 	s \trans{w_3}s, p\trans{w_1[:k]w_2} p \trans{w_3} p, q 
 	\trans{w_1[:k]w_2} q 
 	\trans{w_3}q$ satisfy the following: the sequence of 
 	transitions used while reading the first $k$ letters of $w_2$ is 
 	the same sequence used for reading the first $k$ letters of $w_3$.
 	
 	 ($\Leftarrow$) 
	We will construct a word $w^i = w_1 \cdot 
 	 w_2^0 \cdots w_2^i$ for each $i \ge 0$ such that $\initS \trans{w^i} s$, $\initpa \trans{w^i} p$ and $\initwa 
 	 \trans{w_2^0} q \trans{w_2^1} \cdots q \trans{w_2^i} q$. We let 
 	 $w^{-1}=w_1$ for convenience.
	 For every 
 	 $i$, there will be a new window starting at $|w_1|+|w_2^0 \cdots 
 	 w_2^i|$. The state $\initwa$ will 
 	 be updated to $q$ after reading $w_2^{i+1}$ and keeps coming 
 	 back to $q$ after reading $w_2^j$ for $j > i+1$. 
	 Therefore, none of these 
 	 windows will close during the processing of $w_i$. 
	 Since $q$ is not a dead state, the stream can be extended 
	 with a suffix to realize arbitrarily many overlapping windows.
	 All $w_2^i$ will be built from $w_2$ 
 	 using the completion property, ensuring that $w_2^{i+1}$ makes 
 	 the automaton behave exactly like $w_2^i$ did. We will do this with 
 	 an inductive construction, for which we need to introduce some 
 	 terminology.
 	 
 	 Suppose $w$ is a stream. Consider the substring of $w$ between 
 	 positions $i-k$ to $i$. We 
 	 would like to capture the constraints put on this substring by some 
 	 transition $\tau$ executed by an automaton at the 
 	 $(i+j)$\textsuperscript{th} position, where $j \in [0,k]$. For that 
 	 transition, the values for lookback variables $x_{-k}, \ldots, x_{-j}$ 
 	 are given by $w[i-k+j], \ldots, w[i]$ respectively. For a transition 
 	 $\tau$ and $j \in [0,k]$, let 
 	 $\Phi(\tau)\restr j$ be the set of all atomic 
 	 formulas $\phi$ occurring in 
 	 the guard of $\tau$ such that only the lookback variables $x_{-k}, 
 	 \ldots, x_{-j}$ 
 	 are used in $\phi$. For such an atomic formula 
 	 $\phi$, let $\phi[\rightarrow j]$ be the formula obtained from 
 	 $\phi$ by replacing every lookback variable $x_{-l}$ by 
 	 $x_{-l+j}$ (this results in the values for lookback variables 
 	 $x_{-k+j}, 
 	 \ldots, x_{0}$ of $\phi[\rightarrow j]$ being given by $w[i-k+j], 
 	 \ldots, w[i]$ respectively). In the run $\initS \trans {w_1} s 
 	 \trans{w_2} s \trans{w_3}s$, let $\tau_i$ be the transition executed 
 	 while reading the $i$\textsuperscript{th} letter of $w_2w_3$. For $i 
 	 \in [0,|w_2|-1]$, the 
 	 constraints satisfied by the substring $w_1 w_2[|w_1|+i-k: 
 	 |w_1|+i]$ for this run is $\cup_{j\in[0,k]}\set{\phi[\rightarrow j] 
 	 \mid \phi \in \Phi(\tau_{i+j}) \restr j, \val_V(w_1 w_2[|w_1|+i-k: 
 	 |w_1|+i]) \models \phi[\rightarrow j]} \cup \cup_{j\in[0,k]}\set{\lnot 
 	 \phi[\rightarrow j] 
 	 \mid \phi \in \Phi(\tau_{i+j}) \restr j, \val_V(w_1 w_2[|w_1|+i-k: 
 	 |w_1|+i]) \not\models \phi[\rightarrow j]}$. Let us call this set 
 	 $\Gamma_i^S$. We similarly define the set $\Gamma_i^P$ for the 
 	 run $\initpa \trans{w_1} p\trans{w_2} p \trans{w_3} p$. In the run 
 	 $q \trans{w_1[:k]w_2} q \trans{w_3}q'$, let 
 	 $\tau_i$ be 
 	 the transition executed 
 	 while reading the $i$\textsuperscript{th} letter of $w_2w_3$. For $i 
 	 \in [0,|w_2|-1]$, the 
 	 constraints satisfied by the substring $(w_1[:k] w_2)[i: 
 	 i+k]$ for this run is $\cup_{j\in[0,k]}\set{\phi[\rightarrow j] 
 	 	\mid \phi \in \Phi(\tau_{i+j}) \restr j, \val_V((w_1[:k] w_2)[i: 
 	 	i:k]) \models \phi[\rightarrow j]} \cup 
 	 	\cup_{j\in[0,k]}\set{\lnot 
 	 	\phi[\rightarrow j] 
 	 	\mid \phi \in \Phi(\tau_{i+j}) \restr j, \val_V((w_1[:k]  w_2)[i: 
 	 	i+k]) \not\models \phi[\rightarrow j]}$. Let us call this set 
 	 $\Gamma_i^W$.
 	 
 	 We claim that for every $i \ge 0$, there exists string $w_2^i$ such 
 	 that $|w_2^i|=|w_2|$ and
	for every $j \in 
 	 [0,|w_2|-1]$, $\val_V(w^i[|w^{i-1}|+j -k: 
 	 |w^{i-1}|+j])$ satisfies all the constraints in $\Gamma_j^S \cup 
 	 \Gamma_j^P \cup \Gamma_j^W$. This is sufficient to establish the 
 	 result, since all the automata $S,\pa,\wa$ can repeat the same 
 	 sequence of transitions for $w_2^{i+1}$ as the sequence for 
 	 $w_2^i$. We prove the claim by induction on $i$. For the base 
 	 case 
 	 $i=0$, we set $w_2^0 = w_2$ and the claim is satisfied by 
 	 definition. For the induction step, suppose we have defined up to 
 	 $w_2^i$ as claimed. We define $w_2^{i+1}[j]$ for every $j \in 
 	 [0,|w_2|-1]$ by secondary induction on $j$. 
	 
	 For the base case, 
 	 $j=0$, by the primary induction hypothesis, 
 	 $\val_V(w^i[:k+1])$ satisfies all the constraints in 
 	 $\Gamma_{|w_2|-1}^S \cup \Gamma_{|w_2|-1}^P \cup 
 	 \Gamma_{|w_2|-1}^W$. Let $V' = V\setminus \set{x_0}$. Since the 
 	 first $k$ transitions for $w_3$ are same as the first $k$ transitions 
 	 for 
 	 $w_2$, we infer that $\val_{V'}(w^i[:k])$ satisfies those 
 	 formulas in $\Gamma_{0}^S \cup \Gamma_{0}^P \cup 
 	 \Gamma_{0}^W$ that don't use $x_0$. By the completion property, 
 	 $\val_{V'}(w^i[:k])$ can be extended to include a 
 	 valuation for $x_0$ so that the resulting valuation satisfies all the 
 	 formulas in $\Gamma_{0}^S \cup \Gamma_{0}^P \cup 
 	 \Gamma_{0}^W$. This new valuation for $x_0$ is the value we set 
 	 for $w_2^{i+1}[0]$. The induction step for $j+1$ is similar. This 
 	 completes the induction step and hence establishes the result.
	 \qed
 \end{proof}

 \para{Deciding the Window Overlap Problem} 
The characterization of~\Cref{lem:wop-characterization} 
can now be used to obtain a decision procedure 
for the window overlap problem, provided the alphabet 
theory itself is decidable. This involves checking that there exist words 
$w_1,w_2,w_3$ as claimed above, which can be done using symbolic 
models, 
which decompose the problem into an automata-theoretic  
problem over finite alphabets and satisfiability of formulas in 
the alphabet theory. 

Decidability for 
window overlap problem for decidable alphabet theories with
the completion property can be proven by using techniques similar to those of 
\cite[Theorem 4.4]{AutomataApproachCLTL} (attributed originally to 
\cite{BalbianiCondotta}). This, when converted to terminology used in 
this paper, states that \textit{checking non-emptiness of the language of a 
$\ksla$ can be reduced to checking satisfiability of a finite set 
of formulas in the alphabet theory, provided it is simple and has the 
completion property.}

\section{Stream Processor for WEX with Panes}
\label{sec:processor}
In this section, we present the pseudocode of a stream processor 
that takes a 
WEX and incoming streaming data and produces 
windows as specified in the expressions. For modeling purposes, we 
treat a stream as an infinite string in $\Sigma^\omega$. The data 
present in a window is usually processed to produce an aggregate 
value, such as the average of some field, sum of all entries in a window,
etc. If some positions of the stream belong to multiple windows, 
performing the same 
computation multiple times could be inefficient. One way to avoid this is to 
subdivide windows into \textit{panes} \cite{NoPaneNoGain} and then 
aggregate the values of those panes that make up a window. For 
example, suppose the average of a numerical field is to be computed 
for every window. If within the span of a window, other windows 
start, then the window is subdivided into panes as shown in~\Cref{fig:paning-illustration}.

\begin{figure*}[htb]
\begin{center}
\resizebox{0.75\linewidth}{!}{ 
\begin{tikzpicture}[shorten >=2pt,->, scale=1]
\def\recsize{0.5}
  \tikzstyle{vertex}=[rectangle,fill=black!25, minimum size=12pt, inner sep=0pt, minimum width = \recsize cm, minimum height = \recsize cm, opacity=0.9]
  \tikzstyle{enclose}=[rectangle, thick, draw=blue!90, minimum size=12pt, inner sep=0pt, minimum height = \recsize cm, opacity=0.9] 
    
  \foreach \wtext/\yval/\drop/\panesize/\windowstart/\windowsize in {1/0/0/2/0/0, 2/-\recsize/2/3/0/5, 3/-\recsize*2/5/3/2/6, 4/-\recsize*3/8/5/8/5}
  {

      \foreach \text/\name/\x in {a_0/a_0/0, a_1/a_1/1, a_2/a_2/2, a_3/a_3/3, a_4/a_4/4, a_5/a_5/5, a_6/a_6/6, a_7/a_7/7, a_8/a_8/8, a_9/a_9/9, {a_{10}}/{a_10}/10, {a_{11}}/{a_11}/11, \ldots/{a_{12}}/12}
        {
            \ifthenelse{\NOT \x < \drop}
            {\node[vertex] (G-\name) at (\x/2,\yval) {$\text$}}
            {\node[vertex, opacity=0.1] (G-\name) at (\x/2,\yval) {$\text$}};
        }
        \pgfmathsetmacro\panew{(\panesize)/2)};
        \pgfmathsetmacro\shift{(\drop*\recsize)+(\panew*\recsize)-(\recsize/2)};
        \ifthenelse{\panesize > 0}
        {\node[enclose, minimum width=\panew cm] (a) at (\shift,0) {}}
        {};

        \pgfmathsetmacro\winw{(\windowsize)/2)};
        \pgfmathsetmacro\wshift{(\windowstart*\recsize)+(\winw*\recsize)-(\recsize/2)};
        \ifthenelse{\windowsize > 0}
          {\node[enclose, minimum width=\winw cm, dashed, thick, draw=orange!100,, opacity=0.9] (\wtext) at (\wshift,\yval) {}}
          {};
   }

    \foreach \xstart/\yend/\ystart/\text in {
    0/0/0.7/begin/, 
    2/1/0.7/begin/,
    5/2/0.7/end/,
    8/3/0.7/end \& begin/}
    {
        \pgfmathsetmacro\xpos{\xstart/2 - \recsize/2};      
        \pgfmathsetmacro\yendpos{((\recsize/2) - (\yend*\recsize))};
        \draw[draw=red!70, thick, opacity=0.9] (\xpos, \ystart) node[above, text centered, minimum height=0.5cm,  ]{\text} -> (\xpos, \recsize/2);
    }
    \node[align=right] at (-2.2,1) {Window borders $\rightarrow$};
\matrix [draw,below left] at (current bounding box.west) {
  \node [enclose, label=right:Pane] {}; \\
  \node [] {};\\
  \node [enclose, thick, dashed, draw=orange!100, opacity=0.9, label=right:Window] {}; \\
};

\end{tikzpicture}
}
\end{center}
\caption{Paning: Start a new pane at each instant when a window begins 
or ends}
\label{fig:paning-illustration}
\end{figure*}
For each pane, the average and number of entries in the tuple is 
computed and stored. When the window ends, these can be used to 
compute the average of the whole window. The exact 
computation to be performed in panes is application dependent. We 
assume 
that a class is provided to do that computation. In our 
processor, we 
ensure that the class methods are called at the correct positions in the 
stream.

The pseudocode of the processor is shown in 
Algorithm~\ref{algo:streamProcessor}. For 
simplicity of presentation, we show the processor for one pair of 
expressions $(r,r')$. We further assume that the pair has been 
converted to a pair $(\pa,\wa)$ of prefix and window $\ksla$, both 
deterministic.
It is routine to extend the processor to handle multiple pairs. 
The main idea is that we run the prefix automaton, and whenever it
reaches a final state, we spawn a new copy of window automaton. A new window
is produced when one of the spawned window automaton reaches its final
state. We also perform \textit{garbage collection} by removing window
automaton which are in dead state.

The variable 
$\prefState$ stores the current state of the prefix automaton $\pa$. 
The variable $\windowStartIndices$ stores a mapping 
$\windowStartIndices: Q \to 2^\Nat$, which tracks multiple copies of 
the window automaton $\wa$, as explained next. A copy of $\wa$ is 
started at any position of the input stream that is potentially a start 
position of a window. If such a copy started at position $x$ is 
currently in state $q$, then $x \in \windowStartIndices(q)$. In other 
words, all the copies of $\wa$ that are currently in state $q$ are 
tracked by storing their starting positions in $\windowStartIndices(q)$.

Simultaneously, we generate panes with the following strategy:
\begin{itemize}
  \item Start a new pane at every possible beginning of a new window, i.e., whenever
  the prefix automaton reaches an accepting state.
  \item End the current pane whenever a new pane must be created because a window begins or ends
\end{itemize}

The correctness of the algorithm, formalized in the following results, follows directly
from algorithm's execution.

\begin{algorithm*}
\SetKwInOut{Input}{Input}\SetKwInOut{Output}{output}
\Input{A pair of $\ksla$s $\mathcal{R} = (\pa, \wa)$,  
channels $I,O$ for input, output}
\KwResult{Stream from $I$ decomposed into windows according to
$\mathcal{R}$ and output into $O$}
\SetKwInOut{Nomenclature}{Nomenclature}
\Nomenclature{$Q,Q'$: set of states of $\pa,\wa$}

\caption{Stream processor to extract windows from a stream}
\label{algo:streamProcessor}
\SetKwProg{Fn}{Function}{ is}{end}

\Fn(){main}
{
$\prefState := q_0$ \tcc*{$q_0$ is the initial state of $\pa$}
$\windowStartIndices(q') := \emptyset$ for all $q' \in Q'$\;
$\wordblock := \text{read first $k$ elements of $I$}$\;
$\textit{x} := k-1$\; %
$p := \textbf{new } \textit{Pane}(StartIndex = 0, EndIndex = k-1)$\; %

Update value of $p$ to the aggregate of $\wordblock$\;
$P := \{p\}$ \tcc*{ $P$ is the set of Panes}

\ShowLnLabel{processor:mainLoop}\While{input stream $I$ is live} 
    {
        $\sigma := \text{next item in } I$\; 
        $\wordblock := \wordblock[-k+1:0]\cdot \sigma$\;
        $\startNewPane := $ \textbf{False}\;
        \If{$\prefState \text{ is a final state of 
        } \pa$}
        {
            \ShowLnLabel{processor:prefStateFinal}Add the index $(x+1)$ 
            to $\windowStartIndices(\initwa)$ ($\initwa$ 
            is the initial sate of $\wa$)\;
     
            $\startNewPane := $ \textbf{True}\;
        }
        \ForEach{$q_f \in Q'$ such that $q_f$ is a final state of $\wa$ and \\\hspace{2em}$\windowStartIndices(q_f) \ne \emptyset$}
        {
                \ShowLnLabel{processor:addWindow}Add $(x', x)$ to the 
                output stream $O$ for each index $x' \in 
                \windowStartIndices(q_f)$, along with the aggregate of those 
                panes in $P$ that are between $x'$ and $x$\;
                $\startNewPane := $ \textbf{True}\;
        }
        \ForEach{$q_d \in Q'$ such that $q_d \text{ is a dead state of } \wa$}
	    {
	    	$\windowStartIndices(q_d) := \emptyset$\;
	    }
        \If{$\startNewPane == $ \textbf{True}}
        {
            $p$.EndIndex = $x$;
            $p = new Pane()$;
            $P$.add($p$)\;
        }
        $p$.add\_element($\sigma$) \tcc*{update the current pane with 
        the newly read symbol}
        Let $x_{\min}$ be the minimum of the indices in the range of 
        $\windowStartIndices$\;
        From $P$, delete those panes that end before $x_{\min}$\;
        Update $\prefState$ to new state of $\pa$ by 
        reading $\wordblock$\;
        \ForEach{$\set{q \in Q'}$}
        {
          $Q_{\mathrm{pred}}=\set{q'\in Q' \mid q \text{ is the } 
          \wordblock \text{ successor of } q' \text{ in }\wa}$\;	
          \ShowLnLabel{processor:windowStartUpdate} 
          $\windowStartIndices'(q)= \cup_{q' \in 
          Q_{\mathrm{pred}}}
          \windowStartIndices(q')$\;
        }
        \ShowLnLabel{processor:windowStartAssign} $\windowStartIndices 
        := \windowStartIndices'$ \tcc*{ 
        simultaneous update}
        $x \pluseq 1$\;
    }
}
\end{algorithm*}

\begin{lemma}[Main loop invariant]
	\label{lem:mainLoopInvariant}
The following hold at the start of every iteration of the main loop of the 
processor (line~\ref{processor:mainLoop}), where $n$ is the position of 
the last 
symbol read from the input channel $I$ :
\begin{enumerate}
    \item\label{invariant:prefstate} $\prefState$ stores the state of PA 
    after reading $w[0:n]$.
    \item\label{invariant:window_start} 
    For every position $i < n$ such that $\pa$ reaches a  final state on 
    reading $w[0:i]$, if $\initwa \trans{w[i+1:n]} q$ and $q$ is not a 
    dead-state, then $i+1 \in \windowStartIndices(q)$. 
\end{enumerate}
\end{lemma}
\begin{proof}
We prove this by induction on $n$. Before starting to read symbols 
from $I$, \ref{invariant:prefstate} and 
\ref{invariant:window_start} hold due to the initializations done 
before 
entering the main loop at line~\ref{processor:mainLoop}. Assuming 
that 
they hold at the start of the loop after reading $n$ symbols, we will 
prove that after 
execution of the loop, the invariants would still hold.

Inside the loop, if $\prefState$ is final in $\pa$, then we add the next 
position to $\windowStartIndices(\initwa)$ in 
line~\ref{processor:prefStateFinal}. The states already in 
$\windowStartIndices$ are updated according to the transition 
relation in lines~\ref{processor:windowStartUpdate} and 
\ref{processor:windowStartAssign}. Hence, the invariants continue to 
hold after updating the position variable $x$ in the line following  
\ref{processor:windowStartAssign}.
\qed
\end{proof}

\begin{corollary}
	\label{cor:processorCorrectness}
The processor produced by the Algorithm \ref{algo:streamProcessor}, given the 
input pair $(PA, WA)$ and the stream $w$, will add $(x+1,y)$ to the 
output stream iff $w[0:x] \in L(PA)$ and $w[x+1:y] \in L(WA)$.
\end{corollary}

\begin{proof}
Suppose $(x,y)$ is added to the output, then using 
Lemma~\ref{lem:mainLoopInvariant} (main loop invariant), we get 
$w[0:x-1] \in L(PA)$ and $w[x:y] \in L(WA)$. 
Conversely, suppose $w[0:x-1] \in L(PA)$ and $w[x:y] \in L(WA)$. 
Then 
Lemma~\ref{lem:mainLoopInvariant} implies that PrefixState 
stores the state of $PA$ after reading $w[0:x-1]$, which must be 
a final state, and hence we would have added $x$ to 
$\windowStartIndices(\initwa)$ in 
line~\ref{processor:prefStateFinal}.  
Now, on reading up to the position $y$, we will have
$x \in WindowStateIndices(q)$ such that $q$ is the state of WA after 
reading the word $w[x:y]$, as implied by 
Lemma~\ref{lem:mainLoopInvariant}.

Since $w[x:y] \in L(WA)$, $q$ would be a final state in $WA$, and 
hence the processor must add $(x,y)$ in the output stream $O$ in 
line~\ref{processor:addWindow}.
\qed
\end{proof}

\para{Memory Requirement of Stream Processor}
In the processor produced by the 
Algorithm~\ref{algo:streamProcessor}, $\windowStartIndices$ and $P$ 
are variables that store starting positions of windows and panes 
respectively. 
If the number of starting positions or the number of panes 
is too large, the memory required to store them will also be large.
It is decidable to check whether the memory requirement is unbounded
for a given WEX and input specifier,
when the alphabet theory has the completion property, 
as this problem is equivalent to the
window overlap problem.

\section{Conclusion}
A common feature is to process data from multiple 
streams simultaneously, to perform computations. 
An open problem is to 
integrate this feature into the formal framework of WEX.
Another important direction is to
characterize the fragments with 
decidable window overlap problem beyond the completion property. 
It would also be valuable to study the 
complexity of this problem, the heuristics and algorithms 
that can perform well in practice.
Finally, an orthogonal direction is to 
investigate query optimization 
over WEX, i.e., finding an equivalent WEX with lower memory or computational
requirements. 

\para{Acknowledgment} We thank the anonymous reviewers for their 
critical reading of our paper and valuable feedback,
which greatly
improved its quality.

\bibliographystyle{splncs04}
\bibliography{biblio.bib}

\end{document}